\documentclass[12pt,prl,preprint,showpacs,showkeys]{revtex4}
\usepackage{amsfonts}

\usepackage{graphicx}
\usepackage{dcolumn}
\usepackage{bm}
\include{biblio}

\begin{document}

\title{All-optical control in metal nanocomposites due to a reversible transition between the local-field-enhancement and a local-field-depression upon irradiation by ultrashort control-pulses of light}

\author{Song-Jin Im, Gum-Song Ho}
\affiliation{Department of Physics, Kim Il Sung University, Daesong District, Pyongyang, DPR Korea}

\begin{abstract}
We theoretically study on non-perturbative effective nonlinear responses of metal nanocomposites based on the intrinsic third-order nonlinear response of metal nanoparticles. The large intrinsic third-order nonlinear susceptibility of metal nanoparticles and an irradiation by ultrashort control pulse of light with a sufficiently high peak intensity and moderate fluence can induce a local-field-depression and a saturated plasmon-bleaching in metal nanoparticles. If the control pulse is on, the metal nanocomposites can behave like a dielectric due to the local-field-depression, while if the control pulse is off, the metal nanocomposites can behave like a metal showing a high absorption due to the local-field-enhancement at the plasmon-resonance. This phenomenon can be applied to an ultrafast and remote control of light in metal nanocomposites.
\end{abstract}



\maketitle

\section{1.	Introduction}
The nonlinear properties of metal nanoparticles have been extensively studied for applications in optics, medicine and biology \cite%
{1}. Metal nanocomposites have great perspectives as the nonlinear optical materials because of the large intrinsic nonlinearities of metal nanoparticles \cite%
{2,3,4,5} and the local-field-enhancement near the plasmon-resonance \cite%
{6} resulting in significantly enhanced effective nonlinearities.

Negative nonlinear absorptions which is called the saturated absorptions in metal nanoparticles were experimentally observed \cite%
{7,8,9,10,11,12,13,14,15,16}  and interpreted by the plasmon-bleaching related with the intrinsic electron dynamics in the metal nanoparticles \cite%
{12,13,17}. For higher intensities positive nonlinear absorptions which is called the reverse-saturated absorptions were observed \cite%
{9,11,12,16,17} and attributed to different processes such as the electron-ejection \cite%
{12,19} and multi-photon absorption connected with interband transition \cite%
{10,12,18} in metal nanoparticles. Some studies reported a decrease of negative nonlinear absorption coefficient for higher intensities and attributed it to the reverse-saturated absorption effects starting to play an important role in the overall nonlinear response  \cite%
{10,12}. However, the reverse-saturated absorption effects are overlapping with other effects and strongly dependant on laser characteristics such as fluence, pulse width and repetition rate, thus the quantitative considerations for individually extracted effects are very difficult and complicated. Different experimental results were observed even at the same intensities and wavelengths and the interpretation of the experimental results are controversial \cite%
{12,18}.

In theoretical works the third-order and fifth-order effective nonlinear responses of metal nanocomposites has been studied by approaches such as so-called generalized Maxwell-Garnett model \cite%
{20} and T-matrix method \cite%
{21}, respectively, which is valid for relatively low intensities below 1MW/cm$^{2}$. In \cite%
{22}, the intensity-dependent change of the permittivity of the metal nanoparticles was taken into account by the self-consistent way mainly for intensities  which is about the saturation intensity (defined as the intensity at which the linear loss is reduced by a factor of 2) for which the local-field-factor was reduced, but still larger than unity showing the local-field-enhancement. They predicted that the metal nanocompoites could be used as saturable absorbers with low saturation intensities in the MW/cm$^{2}$ range. Although some calculations for intensities in the GW/cm$^{2}$ range \cite%
{22} were obtained, the results were not interpreted in particular.

In this paper we theoretically study in more detail non-perturbative effective nonlinear responses of metal nanocomposites for higher intensities in the GW/cm$^{2}$ range for which the local fields can be rather depressed than enhanced and the plasmon-bleaching, which causes the saturated absorptions, can be saturated. Based on the Maxwell-Garnett formalism and the intrinsic third-order nonlinear response of metal nanoparticles, disregarding the reverse-saturated absorption effects, we predict that the negative nonlinear absorption coefficient decreases for higher intensities. We predict a transition between the local-field-enhancement and the local-field-depression upon irradiation by ultrashort control-pulses of light which can be used to a remote control of light in metal nanocomposites with a control-light.

\section{2.	Non-perturbative effective nonlinear response of metal nanocomposites}

Effects of morphologies and compositions of nanocomposites on their effective linear optical properties can be described by Maxwell-Garnett formalism. Based on the intrinsic third-order nonlinear response of metal nanoparticles, so-called Generalized Maxwell-Garnett (GMG) model gives the expression for the third-order effective nonlinear response of metal nanocomposites \cite%
{20}.
 
\begin{eqnarray}
\varepsilon_{eff}\approx\varepsilon_{GMG}=\varepsilon_{0}+\nonumber\\
f\chi_{m}^{(3)}\left|\frac{3\varepsilon_{d}}{\varepsilon_{m}+2\varepsilon_{d}}\right|^{2}\left(\frac{3\varepsilon_{d}}{\varepsilon_{m}+2\varepsilon_{d}}\right)^{2}\left|E_{0}\right|^{2}
\label{eq:1}
\end{eqnarray}                          
Here $\varepsilon_{0}$ represents the effective linear permittivity of the nanocomposites and $\varepsilon_{m}$ and $\varepsilon_{d}$ the permittivity of the metal nanoparticles and the dielectric host medium, respectively. $f$ represents the fill-factor, $\chi_{m}^{(3)}$ the intrinsic third-order nonlinear susceptibility of the metal nanoparticles and  $E_{0}$ the incident electric field strength. The part inside the parenthesis corresponds to the local field factor $x$. The GMG model gives insights into the enhanced effective nonlinearity and saturated absorption of the nanocomposites at the plasmon-resonance. However, for a sufficiently high peak intensity of light, the GMG model would become unresonable because the local field factor can have a great change due to a significant nonlinear shift of the permittivity of the metal nanoparticles at the plasmon-resonsnce. Although higher-order models were suggested \cite%
{21}, these models also can not describe non-perturbative effective nonlinear responses of the nanocomposites for the high peak intensity of light.

Combining the intrinsic third-order nonlinear response of the metal nanoparticles and the Maxwell-Garnett formalism describing effects of the morphology and composition of the nanocomposites, the effective permittivity of the nanocomposites can be expressed by the equation (\ref{eq:2}).

 \begin{eqnarray}
\varepsilon_{eff}=\varepsilon_{d}\frac{1+2\left(1-x\right)f}{1-\left(1-x\right)},\nonumber\\
\varepsilon_{m}=\varepsilon_{m0}+\chi_{m}^{(3)}\left|xE_{0}\right|^{2},\nonumber\\
x=\frac{3\varepsilon_{d}}{\varepsilon_{m}+2\varepsilon_{d}}.
\label{eq:2}
\end{eqnarray}                                                  
The first equation of the expression (\ref{eq:2}), which is the Maxwell-Garnett equation, expressed by the local field factor $x$ , is intuitive because the local field factor is the most important parameter determining the effect of the morphology and composition of the nanocomposites on their effective optical properties. $\varepsilon_{m0}$  is the linear permittivity of the metal nanoparticles. It is well worthy to be noted that in the limit of low intensity of light and small fill-factor, the equation (\ref{eq:2}) gives the GMG model. We solve the equation by the self-consistent way.

\begin{figure}
\includegraphics[width=0.8\textwidth]{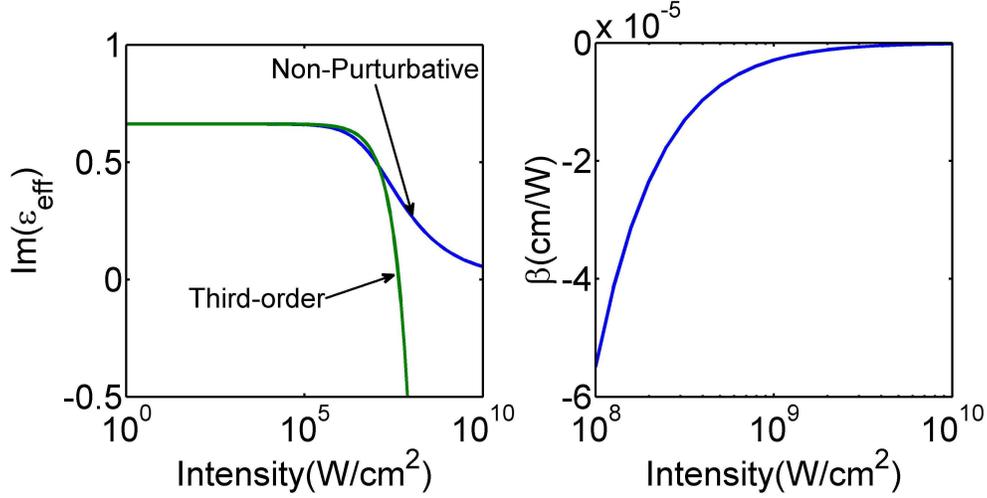}
\caption{(Color online) Non-perturbative effective nonlinear responses of the fused silica doped with silver nanoparticles with the fill-factor f=0.05 at the wavelength $\lambda =450 nm$ according to the incident intensity of light. (a) The imaginary part of the effective permittivity of the nanocomposite calculated by the equation (2) and the GMG model. (b) The nonlinear absorption coefficient   $\beta(I)=\Delta\alpha(I)/\Delta I$ calculated by the equation (2).}
\label{fig:1}
\end{figure}

Fig. \ref{fig:1} shows non-perturbative effective nonlinear responses of the fused silica doped with silver nanoparticles with the fill-factor f=0.05 at the wavelength $\lambda =450 nm$ according to the incident intensity of light. The size of the silver nanoparticles is assumed to be about 10 nm which is between the mean free path of the electrons and the skin depth. Electromagnetic interaction between the metal nanoparticles can be ignored by taking the small fill-factor below 0.1 \cite%
{24}. We use the experimental data of the permittivity of silver  \cite%
{25} and take the intrinsic third-order nonlinear susceptibility  $\chi_{m}^{(3)}=(-6.3+1.9i)\times10^{-16}m^{2}/V^{2}$ \cite%
{26}. In Fig. \ref{fig:1}(a), the imaginary part of the effective permittivity of the nanocomposites calculated by the GMG model deviates from that calculated by the equation (2) for intensities higher than 10MW/cm$^{2}$ and even has negative values showing a non-physical amplification. We can know the effective nonlinear response of the nanocomposite becomes non-perturbative for the high intensities. In Fig. \ref{fig:1}(b), the nonlinear absorption coefficient $\beta(I)=\Delta\alpha(I)/\Delta I$  has negative values, showing the saturated absorption due to the plasmon-bleaching, and decrease for higher intensities, showing a saturated plasmon-bleaching. In Ref.  \cite%
{10,12} the decrease of negative nonlinear absorption coefficient was experimentally observed and attributed to positive nonlinear absorption effects such as the electron-ejection in metal nanoparticles. We note that the intrinsic third-order nonlinear response of metal nanoparticles disregarding the positive nonlinear absorption effects and an effect of the morphology and composition of the nanocomposite can also provide the decrease of negative nonlinear absorption coefficient.

\begin{figure}
\includegraphics[width=0.45\textwidth]{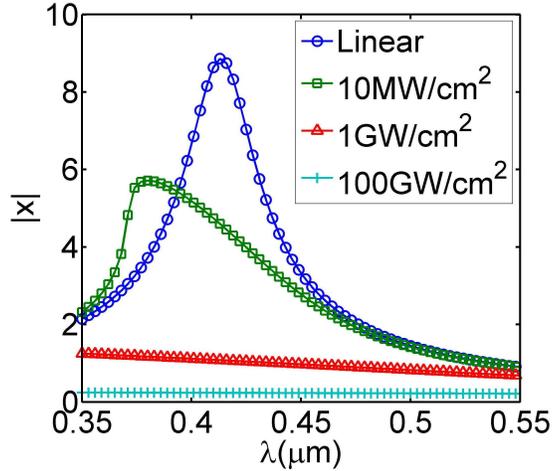}
\caption{(Color online) The absolute value of the local field factor $|x|$ according to the wavelength for different incident intensities in the fused silica doped with silver nanoparticles with the fill-factor f=0.01 calculated by the equation (2).}
\label{fig:2}
\end{figure}

The Fig. \ref{fig:2} shows the local field factor $|x|$ according to the wavelength for different incident intensities in the silver-doped fused silica with the fill-factor f=0.01 calculated by the equation (\ref{eq:2}). In the cases of the low incident intensities, we can see the plasmon-resonance and the local-field-enhancement. In the case of the incident intensity of 10MW/cm$^{2}$, one can see the blue-shift of the plasmon-resonance. For the incident intensities in the range of GW/cm$^{2}$, one can see the saturated plasmon-bleaching and the local-field-depression. The plasmon-resonance and the local-field-enhancement in the metal nanoparticles induce a high absorption in the nanocomposites, however in contrast the saturated plasmon-bleaching and the local-field-depression can induce a transparency in them.

\section{3.	All-optical control in metal nanocomposites}

\begin{figure}
\includegraphics[width=0.8\textwidth]{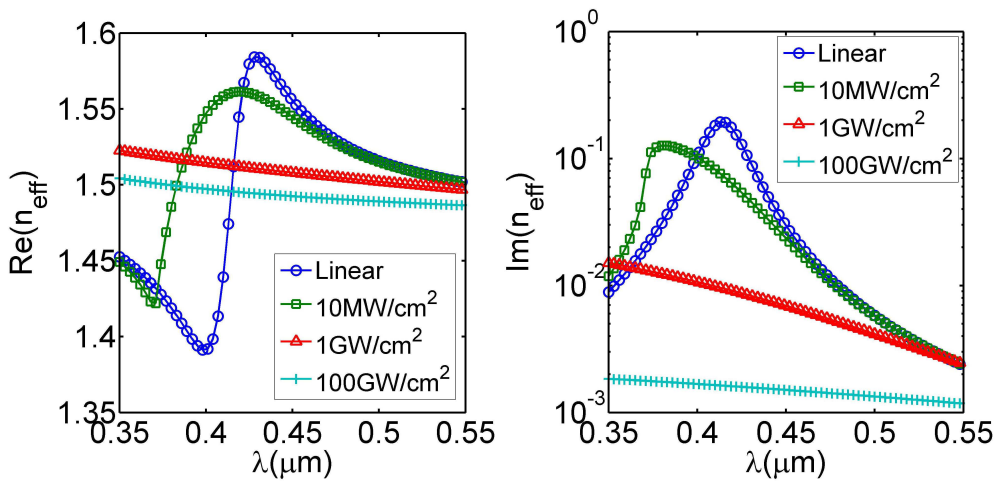}
\caption{(Color online) Real and imaginary parts of the effective refractive index of the fused silica doped with silver nanoparticles with the fill-factor f=0.01 according to the wavelength for different incident intensities calculated by the equation (2).}
\label{fig:3}
\end{figure}

Fig. \ref{fig:3} shows the real and imaginary parts of the effective refractive index of the fused silica doped with silver nanoparticles with the fill-factor f=0.01 according to the wavelength for different incident intensities calculated by the equation (\ref{eq:2}). One can see the low imaginary parts of the effective refractive indices for the high incident intensities showing a low absorption in the nanocomposite as predicted above. The real part of the effective refractive index for the high incident intensities is close to the refractive index of the host dielectric medium. In the ideal case of very high incident intensity the effective permittivity of the metal nanocomposite converges to the value which is determined by the kind of the host dielectric medium and the fill-factor and is independent of the property of the metal.

 \begin{eqnarray}
\varepsilon_{eff,ideal}=\varepsilon_{d}\frac{1+2f}{1-f}
\label{eq:3}
\end{eqnarray} 

\begin{figure}
\includegraphics[width=0.8\textwidth]{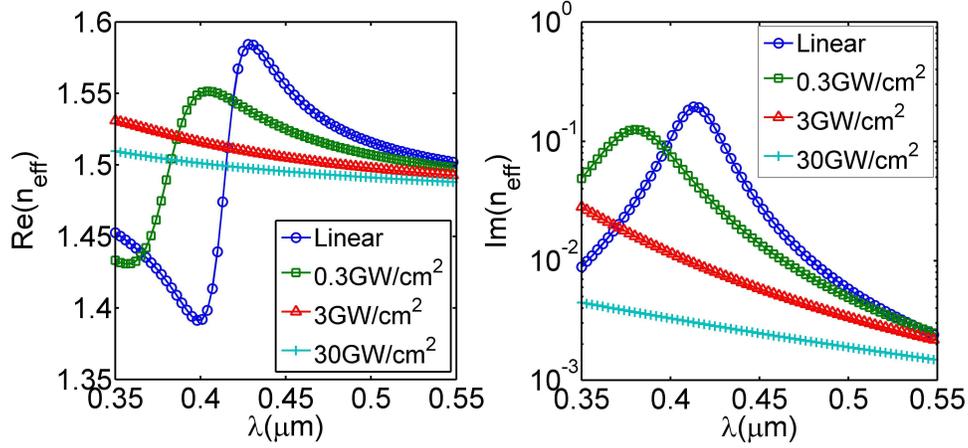}
\caption{(Color online) Real and imaginary parts of the effective refractive index of the fused silica doped with silver nanoparticles with the fill-factor f=0.01 according to the wavelength of the signal-light for different incident intensities of the control-light at the wavelength $\lambda_{c} =500 nm$ calculated by the equation (4).}
\label{fig:4}
\end{figure}

The effective optical properties of nanocomposites can be remotely controlled with a control-light utilizing the cross-nonlinearity in the metal nanoparticles. The cross-nonlinear susceptibility of metal nanoparticles $\chi_{m}^{(3)}(\omega_{s};\omega_{c},-\omega_{c},\omega_{s})$ is predicted to be large like the self-nonlinear susceptibility $\chi_{m}^{(3)}(\omega_{s};\omega_{s},-\omega_{s},\omega_{s})$ , while $\chi_{m}^{(3)}(3\omega_{s};\omega_{s},\omega_{s},\omega_{s})$  for the third-harmonic generation is depressed and smaller by several orders of magnitude than that the self-nonlinear susceptibility \cite%
{10,khk_laserphysics_2013}.
Based on the intrinsic third-order nonlinear response of metal nanoparticles and the Maxwell-Garnett formalism, the effective permittivity of metal nanocomposites irradiated by both the signal-light at the frequency $\omega_{s}$  and the control-light at $\omega_{c}$ can be expressed by the equation (\ref{eq:4}).

 \begin{eqnarray}
\varepsilon_{eff}(\omega_{s})=\varepsilon_{d}\frac{1+2(1-x_{s})f}{1-(1-x_{s})f},\nonumber\\
\varepsilon_{eff}(\omega_{c})=\varepsilon_{d}\frac{1+2(1-x_{c})f}{1-(1-x_{c})f},\nonumber\\
\varepsilon_{m}(\omega_{s})=\varepsilon_{m0}(\omega_{s})+\nonumber\\
\chi_{m}^{(3)}(\omega_{s};\omega_{s},-\omega_{s},\omega_{s})\left|x_{s}E_{0}(\omega_{s})\right|^{2}+\nonumber\\
\chi_{m}^{(3)}(\omega_{s};\omega_{c},-\omega_{c},\omega_{s})\left|x_{c}E_{0}(\omega_{c})\right|^{2},\nonumber\\
\varepsilon_{m}(\omega_{c})=\varepsilon_{m0}(\omega_{c})+\nonumber\\
\chi_{m}^{(3)}(\omega_{c};\omega_{s},-\omega_{s},\omega_{c})\left|x_{s}E_{0}(\omega_{s})\right|^{2}+\nonumber\\
\chi_{m}^{(3)}(\omega_{c};\omega_{c},-\omega_{c},\omega_{c})\left|x_{c}E_{0}(\omega_{c})\right|^{2},\nonumber\\
x_{s}=\frac{3\varepsilon_{d}}{\varepsilon_{m}(\omega_{s})+2\varepsilon_{d}},x_{c}=\frac{3\varepsilon_{d}}{\varepsilon_{m}(\omega_{c})+2\varepsilon_{d}}.
\label{eq:4}
\end{eqnarray} 
Here the subscript s and c represents the signal-light and control-light, respectively. Fig. \ref{fig:4} shows the real and imaginary parts of the effective refractive index of the fused silica doped with silver nanoparticles with the fill-factor f=0.01 according to the wavelength of the signal-light for different incident intensities of the control-light at the wavelength $\lambda_{c} =500 nm$  calculated by the equation (\ref{eq:4}), here we assume that the control-light is far more intensive than the signal-light. We take the cross-nonlinear susceptibility $\chi_{m}^{(3)}(\omega_{s};\omega_{c},-\omega_{c},\omega_{s})$  and the self-nonlinear susceptibility $\chi_{m}^{(3)}(\omega_{s};\omega_{s},-\omega_{s},\omega_{s})$  same as the intrinsic third-order nonlinear susceptibility used in Fig. \ref{fig:1}. One can see that the effective optical properties of nanocomposite can be controlled with an intensive control-light. The nanocomposite irradiated by the control-light can behave like a dielectric showing a low absorption, while in the absent of the control-light can behave like a metal showing a high absorption at the plasmon-resonance.

\begin{figure}
\includegraphics[width=0.45\textwidth]{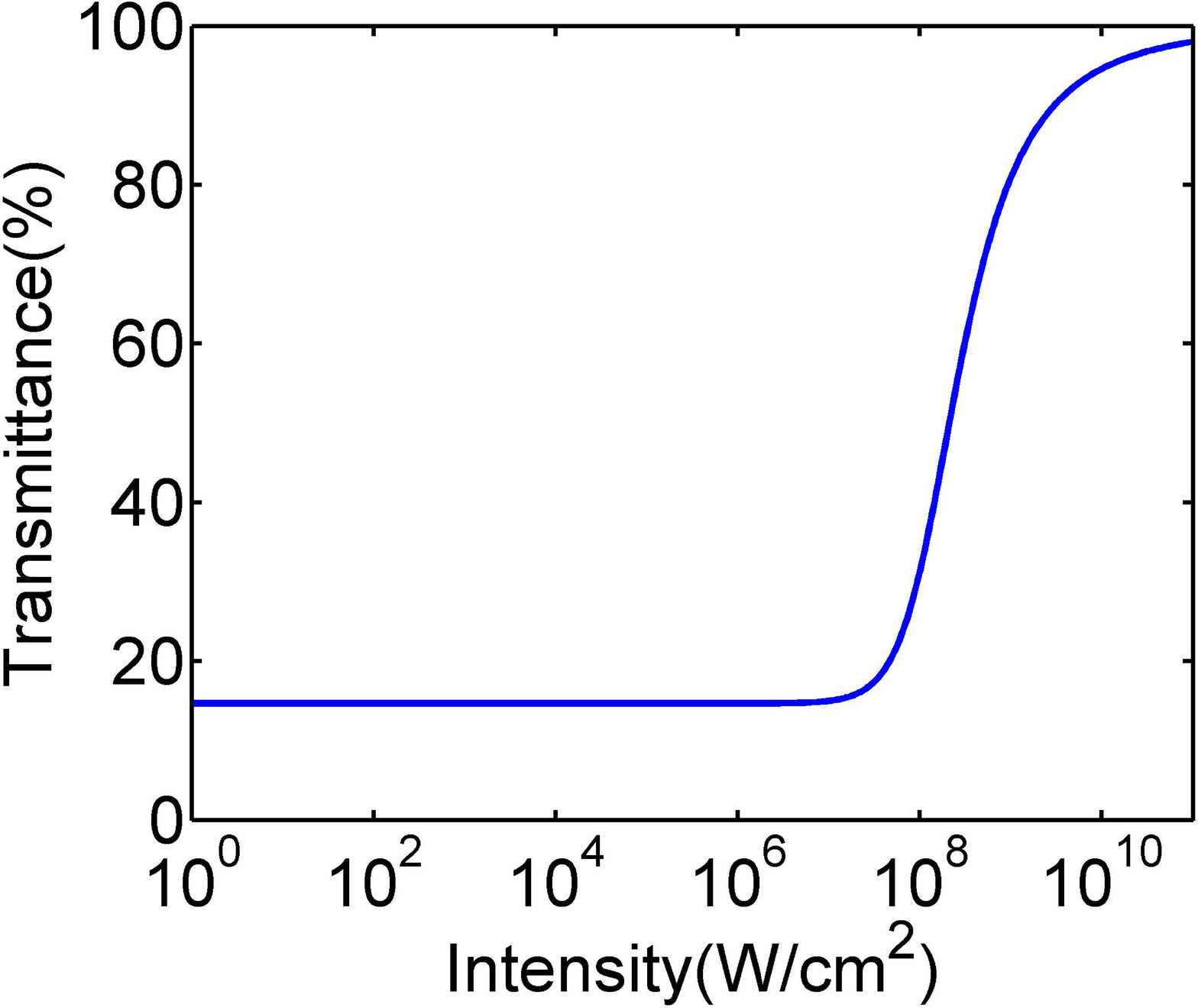}
\caption{(Color online) Transmittance of 70 nm-layer of the fused silica with silver nanoparticles with the fill-factor f=0.05 at the linear plasmon-resonance $\lambda_{s} =414 nm$ according to the peak intensity of the control-light at the wavelength $\lambda_{c} =500 nm$.}
\label{fig:5}
\end{figure}

Fig. 5 shows the transmittance of 70 nm-layer of the fused silica doped with silver nanoparticles with the fill-factor f=0.05 at the linear plasmon-resonance  $\lambda_{s} =414 nm$ according to the peak intensity of the control-light at the wavelength $\lambda_{c} =500 nm$. By sending a control pulse with the high peak intensity in the GW/cm$^{2}$ range together with a synchronized signal pulse train or continuous wave at the linear plasmon-resonance, all-optical switching can be implemented. If the control pulse is on, the signal has a much higher transmission.

\section{4.	Conclusion}

We theoretically predicted that the intrinsic third-order nonlinear response of metal nanoparticles and an effect of the morphology and composition of nanocomposites can result in the non-purturbative effective nonlinear response of the nanocomposites such as the saturated plasmon-bleaching and the local-field-depression and suggested the all-optical control in nanocomposites due to the transition between the local-field-enhancement and local-field-depression. However, the needed intensities in the GW/cm$^{2}$ range can cause positive nonlinear absorption effects resulting in significant changes of the intrinsic third-order nonlinear response of the metal nanoparticles. We assumed sub-picosecond control-pulses of light with moderate fluence and repetition rate would keep the metal nanocomposites from the effects such as electron-ejection, cumulative heating and optically induced irreversible damages.
The switching time in such a device
will be limited by the inherent cooling time which is on the order of 1 ps \cite%
{khk_laserphysics_2013}.

\end{document}